\PassOptionsToPackage{square,comma,numbers,sort&compress,super}{natbib}  
\documentclass[%
aip,
amsmath,amssymb,nofootinbib,
reprint,%
]{revtex4}
\usepackage{graphicx}
\usepackage{xcolor}
\usepackage{unicode}
\usepackage{array}
\usepackage{amsmath}
\usepackage{caption}
\usepackage{subcaption}
\usepackage{mathtools}
\usepackage{hyperref}
\usepackage{booktabs}
\usepackage{dcolumn}
\usepackage{bm}
\usepackage{setspace}
\usepackage[utf8]{inputenc}
\usepackage[T1]{fontenc}
\usepackage{mathptmx}
\usepackage[page,toc,titletoc,title]{appendix}
\usepackage[makeroom]{cancel}
\usepackage{etoolbox}
\usepackage{verbatim}
\newcommand {\curl} {\nabla \times}
\newcommand {\diver} {\nabla \cdot}
\newcommand*{\balancecolsandclearpage}{%
 \close@column@grid
 \cleardoublepage
 \twocolumngrid
}

\newcolumntype{P}[1]{>{\centering\arraybackslash}p{#1}}

\makeatletter
\def\@email#1#2{%
\endgroup
\patchcmd{\titleblock@produce}
 {\frontmatter@RRAPformat}
 {\frontmatter@RRAPformat{\produce@RRAP{*#1\href{mailto:#2}{#2}}}\frontmatter@RRAPformat}
 {}{}
}%
\appto{\appendix}{%
  \@ifstar{\def\theequation@prefix{A.}}%
          {}%
}
\makeatother

\begin{document}


\title{\LARGE{Self-consistent full MHD coupling of JOREK and STARWALL for advanced plasma free boundary simulation}\\
   \large{On the electromagnetic interaction of a Full MHD plasma with a 3D wall model} }

\author{\textbf{R. Sparago}}
\affiliation{Max Planck Institute for Plasma Physics, Boltzmannstrasse 2, 85748 Garching, Germany}
\affiliation{ITER Organization, 13067 St. Paul Lez Durance Cedex, France}
\author{F.J. Artola}%
\affiliation{ITER Organization, 13067 St. Paul Lez Durance Cedex, France}
\author{M. Hoelzl}
\affiliation{Max Planck Institute for Plasma Physics, Boltzmannstrasse 2, 85748 Garching, Germany}%


\begin{abstract}

\noindent
\rule{0.93\textwidth}{2pt}
\section*{\label{Abstract}}
\hspace{-1em}\textbf{ABSTRACT} 

\medskip
An adequate modelling of the electromagnetic interaction of the plasma with the surrounding conductors is paramount for the correct reproduction of 3D plasma dynamics. Simulations of the latter provide in turn useful predictions regarding the plasma evolution, the related MHD modes leading to disruptions and the electromagnetic forces acting on the vacuum vessel's components when said disruptions occur. The latest modelling efforts with the 3D FEM non-linear JOREK code have been directed towards the eddy current coupling of a reduced magnetohydrodynamic (MHD) plasma model with thin and volumetric wall codes (STARWALL and CARIDDI). In this contribution, we present an eddy current coupling between the full MHD model of JOREK and the STARWALL code; this new coupling scheme describes the full three-dimensional interactions of the plasma with the vacuum region and external conductors, modeled by natural boundary conditions linking the magnetic vector potential $\mathbf{A}$ to the magnetic field $\mathbf{B}$. The consistency of the new coupling scheme is validated via benchmarks for axisymmetric Vertical Displacement Events and multi-harmonics simulations of MHD modes.

\bigskip
\hspace{-1em} \rule{0.93\linewidth}{2pt}
\end{abstract}

\maketitle

\section{\label{sec:intro}\normalsize{An introduction to free boundary plasma scenarios}}

It is widely known that an accurate study of the macroscopic dynamics of a tokamak plasma requires that its interaction with the surrounding electromagnetic (EM) environment be taken into account. This serves the purpose  of "informing" the physical model in use about two main phenomena:
\begin{enumerate}
   \item The plasma is immersed in a \textit{vacuum}, i.e. a domain with a certain magnetic permeability $\mu_0$ and electric permittivity $\epsilon_0$, and a zero electric conductivity $\sigma_0$;
   \item The presence of the tokamak's conducting structures (vacuum vessel, first wall, in-vessel components) that surround the plasma determines a wide range of mutual three-dimensional EM interactions between currents in the plasma and the conductors.
\end{enumerate} 
The above-mentioned points enrich the model by introducing a \textit{free boundary} scenario, where the magnetic field at the plasma edge is not anymore \textit{fixed} in space and time, but rather, it is shaped by the interaction with the external electromagnetic environment.
This more comprehensive modelling setting renders itself fundamental for a more accurate reproduction of the macroscopic plasma dynamics and the forces arising on the tokamak structures during a plasma discharge. \\
Such information is crucial when dealing with \textit{disruptions}, that is, violent events caused by the excessive growth of plasma instabilities, which in turn lead to a rapid loss of the thermal and magnetic energy stored within the plasma.

\medskip
Disruptions are often associated with motion, variation in the current and temperature profiles and changes in the magnetic topology of the plasma; said occurrences determine the induction of eddy currents in the surrounding conductive structures and the exchange of current between the plasma and the wall structures; these currents generate a reaction magnetic field that ends up affecting the plasma evolution itself. Aside from this feedback effect, the wall current density field $\mathbf{J_w} $ associated with a plasma perturbation will interact with the magnetic field $\mathbf{B}$, thus generating an electromechanical force distribution $\mathbf{F_w}$ on the machine's structures. \\
Forces that are brought about by certain magnetohydrodynamic (MHD) modes can be detrimental for future reactors' components; furthermore, because of the frequency of some of these modes, the resulting forces can be \textit{rotating} and therefore constitute a threat for the integrity of conducting structures due to resonant excitation\,\, \color{blue} [\onlinecite{pustovitov2021sideways}].
\color{black}
\begin{table*}
   \centering
   \begin{ruledtabular}
       \begin{tabular}{|P{5cm}|P{6cm}|P{6cm}|}
           
           \textbf{Code suit} & \textbf{Plasma model} & \textbf{Conductive structures model} \\
           \hline
           CarMa0NL\, \color{blue} [\onlinecite{villone2013coupling}] \color{black} & 2D evolutionary equilibrium & 3D volumetric \\ 
           \hline
           NIMROD\, \color{blue} [\onlinecite{sovinec2004nonlinear}] \color{black} & 3D full MHD & 2D* thin \\
           \hline
           M3D\, \color{blue} [\onlinecite{park1999plasma}] \color{black} & 3D full MHD & 2D* thin \\
           \hline
           M3D-C$^1$\, \color{blue} [\onlinecite{ferraro2016multi}] \color{black} & 3D full MHD & 2D* volumetric \\
           \hline
           JOREK-STARWALL\, \color{blue} [\onlinecite{hoelzl2012coupling}]\, \color{blue} [\onlinecite{hoelzl2021jorek}]\, \color{blue} [\onlinecite{huysmans2007mhd}]  \color{black} & 3D reduced / full MHD & 2D* thin \\
           \hline
           JOREK-CARIDDI\, \color{blue} [\onlinecite{isernia2023self}] \color{black} & 3D reduced / full MHD & 3D volumetric
       \end{tabular} \hfill
       \caption{\label{Codes review table}\small{ Overview of presently available simulation tools for the reproduction of free boundary MHD modes. *The 2D notation stands for axisymmetry of the conducting structures, but non-axisymmetric wall currents are allowed by the 3D plasma codes nonetheless. Furthermore, simple 3D wall representations are also possible in STARWALL.}}
   \end{ruledtabular}
\end{table*}

\medskip
Because of the above, tokamak loads ensuing disruptions represent a major concern for the machine's mechanical structure, which needs to be designed appropriately leveraging the predictive capability of simulation codes. \\
To this end, a variety of coupled MHD plasma $\leftrightarrow$ EM wall codes has been developed in the past 30 years, each characterized by different degrees of simplifying assumptions concerning both the plasma and the conducting structures' representation. An overview of these codes can be found in Table \ref{Codes review table}. The CarMa0NL\color{blue}\, \color{blue} [\onlinecite{villone2013coupling}] \color{black} code is an example of simplified plasma model coupled with 3D volumetric conductors: here, the plasma is assumed to be massless, thus resulting in instantaneous reactions to external magnetic perturbations; in addition, the MHD equations are solved in the non-linear axisymmetric perturbed equilibrium limit. With different coupling schemes, the 3D full MHD codes NIMROD\, \color{blue} [\onlinecite{sovinec2004nonlinear}] \color{black} and M3D\, \color{blue} [\onlinecite{park1999plasma}] \color{black} were instead coupled to an axisymmetric thin model of passive conductors, thus providing a more limited representation of the complex current distributions that can arise in these conductors. This limit was relaxed in M3D-C$^1$\, \color{blue} [\onlinecite{ferraro2016multi}] \color{black} with the adoption of an axisymmetric volumetric representation for the conductors, which were included in the standard computational domain; however at the cost of a limited resolution for their representation. \\
The JOREK-STARWALL\, \color{blue} [\onlinecite{hoelzl2012coupling}] \color{black} and JOREK-CARIDDI\, \color{blue} [\onlinecite{isernia2023self}] \color{black} code suits are the only frameworks that are currently featuring a non-linear extended MHD plasma model with a \textit{self-consistent} representation of 3D conducting structures. As will be shown, the coupling strategy is based on the virtual casing principle, also adopted in CarMa0NL. The only difference between STARWALL and CARIDDI stems from the numerical representation of the electromagnetic degrees of freedom in the conducting structures; in STARWALL, the conductors are limited to be thin, whereas CARIDDI is able to solve the electromagnetic problem under investigation for the most general case, that is, thick and non-axysimmetric conductive structures. This makes JOREK-CARIDDI the only code with a self-consistent coupling of a 3D MHD plasma with 3D volumetric conductors. The latter representation becomes fundamental when the skin depth of the magnetic perturbation is comparable to the wall width, which can occur for certain MHD modes, and when the modelling of the path of shared (\textit{halo}) currents between the plasma and the wall structures is required. \\
At present, coupling schemes with the JOREK\, \color{blue} [\onlinecite{hoelzl2021jorek}] \color{black} code have been implemented only for eddy currents; a halo current coupling is currently in preparation. Furthermore, for both STARWALL and CARIDDI, the eddy current coupling has been only carried out for the reduced MHD model of JOREK.
The present contribution aims to improve the fidelity of the last-mentioned code suits even further by performing the coupling of the full MHD model of JOREK, which opens up the possibility to reproduce plasma free boundary dynamics and wall forces with a greater accuracy than before. In particular, the \textit{full MHD coupling} of JOREK and STARWALL is presented, with an eye to a straightforward future extension to JOREK-CARIDDI. \\ 
The motivations to implement the full MHD coupling are several. Firstly, the reduced MHD model of JOREK cannot correctly reproduce some instabilities, even if they
are internal and therefore not coupled with the wall. 
In addition, many plasma physics phenomena (ELMs, runaway electrons, toroidal field
ripples, plasma current spikes, current driven modes) need free boundary to be reproduced, and the model’s accuracy certainly benefits from the extension of the field treatment to three dimensions.

\medskip
The modelling features of the two codes and the relevant free boundary conditions are presented in Sec.\@ \ref{Models}. In Sec.\@ \ref{Coupling scheme}, we detail the full MHD coupling scheme adopted to extend the previous reduced MHD model.\@ In Sec.\@ \ref{Tilted wire} vacuum tests are shown for the extended electromagnetic STARWALL response. In Sec.\@ \ref{Simulations} simulations of axisymmetric $(n=0)$ Vertical Displacement Events (VDEs) and non axisymmetric MHD modes are reproduced for the physical validation of the full MHD JOREK-STARWALL suit.

\section{\normalsize{Modelling framework}} \label{Models}
A mathematical description of the problem of interest is presented in this section with the purpose of setting the necessary ground to illustrate the full MHD extension in Section \ref{Coupling scheme}.

\begin{figure}[h]
   \centering
   \includegraphics[width=0.6\linewidth]{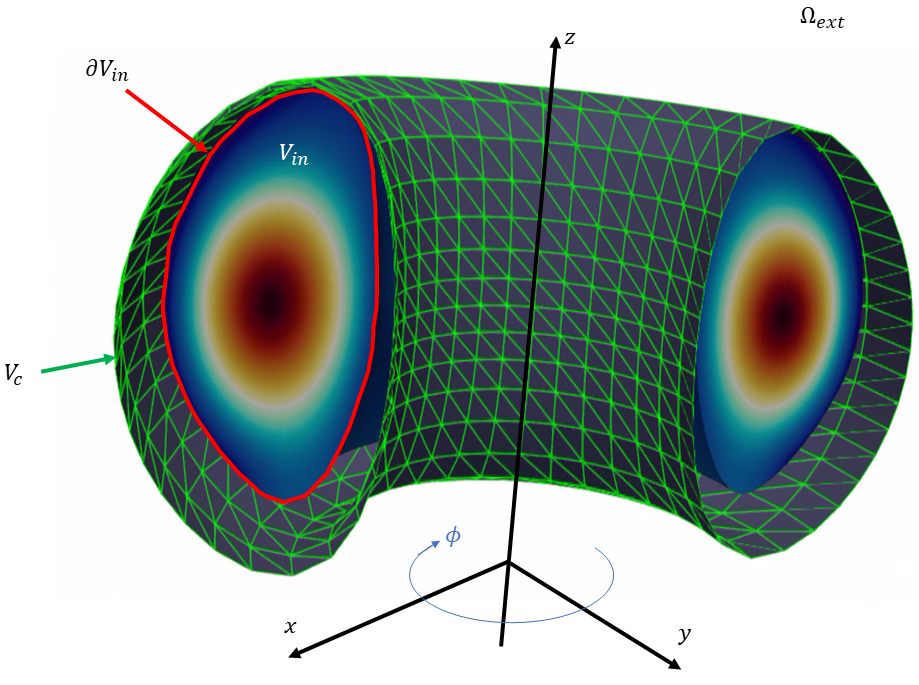}
   \caption{\small{3D visualization of the coordinate system and basic representation of the JOREK and STARWALL domains. For a better distincion, STARWALL's domain is represented in its FEM triangular discretization. In the JOREK frame, the toroidal $\phi$ coordinate increases in the clockwise direction. }}
   \label{fig:JS domain}
\end{figure}

JOREK is a non-linear 3D MHD finite-elements code with a solution domain that encloses the plasma core, the scrape-off layer (SOL) and the divertor region. As such, the MHD domain $V_{in}$ is delimited by a computational boundary $\partial V_{in} $. The conducting structures' domain $V_c$ is immersed in the external space $\Omega_{ext}$. For a visual reference, a representation of the JOREK and STARWALL domains is provided in Fig. \ref{fig:JS domain}.
\\
The coupling hinges on the key principle that the solution to the Magneto Quasi Static (MQS) problem provided by STARWALL be used to inform the corresponding boundary conditions (BCs) in JOREK. \\
The default set of BCs in JOREK are of \textit{Dirichlet} type for all the variables, namely, they are assumed to be fixed in time on the computational boundary. Focusing on magnetic BCs, it should be noted that a Dirichlet BC on
$\textbf{B}$ is physically equivalent to replacing the boundary with a perfectly conducting surface,
where the normal magnetic field is frozen in time (\emph{ideal wall limit}). This condition is equivalent to solving a fixed boundary problem and is of course not suitable to describe a free boundary scenario, where instead the magnetic field evolves in time on the domain's edge: for that, a \textit{Neumann} BC is needed, where the tangential component of the magnetic field $\mathbf{B \times n}$ is provided on the boundary at each time instant. This component is exactly where the response from the external structures enters to include the vacuum and conducting structures' contribution. \\
In particular, the STARWALL code yields the following main result:
\begin{equation} \label{SW yield}
\mathbf{B \times n} = \color{black}f \left({\mathbf{A \times n}}, I_w  \color{black} \right) = \underline{\mathbf{M_{vac}}} \, \mathbf{A \times n} \color{black} + \underline{\mathbf{M}_I} \, I_w 
\end{equation} 
where $f$ is a vector integral operator, $\mathbf{A}$ and $\mathbf{B}$ are the magnetic vector potential and magnetic flux density at discrete points of the JOREK computational boundary, $\mathbf{n}$ is the normal unit vector on each boundary point, $I_w $ contains the wall and coil currents in the conducting structures; $\underline{\mathbf{M_{vac}}}$ and $\underline{\mathbf{M}_I}$ are the matrices computed by STARWALL given the geometry under study. The ${\mathbf{B \times n}}$ term and the wall currents $I_w$ represent the output of the code solving $\Omega_{ext}$, while ${\mathbf{A \times n}}$ is the input from JOREK. It is appropriate to clarify at this point that Equation \eqref{SW yield} is here presented in a general form, accounting for both divergence-free eddy current fields and source-sink shared (\emph{halo}) currents between the plasma and the conducting structures. This work is exclusively dedicated to the full MHD eddy current coupling, while the halo current coupling will be the subject of future work.

\subsection{\label{JOREKside}\normalsize{JOREK-side of the coupling}}
JOREK solves a large family of physics models and is capable of capturing a broad spectrum of phenomena, including parallel/diamagnetic flows, bi-fluid behavior, neutrals, impurities, runaway electrons, radiation physics and kinetic effects. \\
The base system of equations solved in JOREK is the \emph{visco-resistive MHD model}\, \color{blue} [\onlinecite{hoelzl2021jorek}] \color{black}

\begin{align}
   \label{inductioneq} \frac{\partial \mathbf{A}}{\partial t} & = -\mathbf{E}-\nabla \Phi \\ 
   \rho \frac{\partial \mathbf{v}}{\partial t} & = -\rho \mathbf{v} \cdot \nabla \mathbf{v} - \nabla p + \mathbf{J \times B} + \nabla \cdot \underline{\tau} + \mathbf{S_v}   \\
   \frac{\partial \rho}{\partial t} & = -\nabla \cdot \left( \rho \mathbf{v} \right) + \nabla \cdot \left( \underline{D} \nabla \rho \right) + S_{\rho} \\
   \frac{\partial p}{\partial t} & = -\mathbf{v} \cdot \nabla p - \gamma p \nabla \cdot \mathbf{v} + \nabla \cdot \left( \underline{\kappa} \nabla T \right) + \left( \gamma - 1 \right) \underline{\tau} : \nabla \mathbf{v} + S_p \\
  \label{E eq.} \mathbf{E} &= - \mathbf{v \times B} + \eta \mathbf{J}
\end{align}
where $\mathbf{A}$ is the magnetic vector potential, $\mathbf{J}$ is the current density, $\mathbf{B}$ is the magnetic field, $\mathbf{v}$ is the fluid velocity, $\rho$ is the fluid density, $p$ is the pressure, $\mathbf{E}$ is the electric field, $\Phi$ is the scalar electric potential, $\underline{\tau} $ is the viscous stress tensor, $\underline{D} $ is the particle diffusion tensor, $\underline{\kappa}$ is the heat diffusion tensor and $\mathbf{S_v}$, $S_{\rho}$ and $S_p$ are source terms for momentum, density and pressure, $\gamma$ is the ratio of specific heats $ \frac{C_p}{C_v}$, $\eta$ is the electric resistivity. \\
The present set of equations is solved via numerical FEM method applied to a 2D Bézier iso-parametric grid complemented by a Fourier expansion in toroidal direction.\, \color{blue} [\onlinecite{czarny2008bezier}] \color{black}
Starting from this general model, different reasonable assumptions can be introduced, thus leading to the various physics models included in the code. In particular, a clear distinction exists between the two families of reduced and full MHD models that the user can choose to compute the solution. \\
\subsubsection*{Reduced MHD}
The reduced MHD model is based on the following magnetic field ansatz:
\begin{equation} \label{reduced MHD ansatz}
   \mathbf{B} = F_0 \nabla \phi + \nabla \psi \times \nabla \phi
\end{equation} 
where $F_0 = R \, B_{\phi, TF coils}$, with $R$ as the radial coordinate in a cylindrical system, and $\psi$ is the poloidal magnetic flux per radian. Equation \eqref{reduced MHD ansatz} entails two main consequences:
\begin{enumerate}
   \item The only toroidal field contribution $F_0$ is given by the background coils, implying that the plasma cannot compress $B_{\phi} $ during the dynamics, therefore the time-varying part of $\mathbf{A}$ is
   \begin{equation}
       \mathbf{A} = \psi \nabla \phi
   \end{equation}
  where $\phi$ is the toroidal coordinate.
   \item As a consequence of both equation \eqref{E eq.} and the previous statement, the fluid velocity is only comprised of two terms
   \begin{equation}
       \mathbf{v} = R^2 \nabla \phi \times \nabla \left( \Phi / F_0 \right) + v_{\parallel} \mathbf{B}
   \end{equation}
   which stand for the $\mathbf{E} \times \mathbf{B}$ term related to Alfvén waves and the field-aligned flows related to slow magneto-sonic waves. Diamagnetic flows can also be added to this representation. Here, $\Phi$ is the scalar electric potential. Note that $\Phi$ is assigned a Dirichlet-type BC, which entails that currents flowing on the boundary (i.e. SOL currents) see a perfect conductor on the outside; this simplification needs to be removed when the coupling of halo currents with the vacuum and the wall is taken into account, or when sheath boundary conditions are considered.
\end{enumerate}
This ansatz-based reduced MHD model allows to neglect fast magneto-sonic waves, thus reducing the required computational effort. This model is energy conserving and has been shown to capture linear and non-linear dynamics accurately for most instabilities.\color{blue} [\onlinecite{hoelzl2021jorek}] \color{black}

\subsubsection*{Full MHD}
In full MHD models, the $\mathbf{A,B}$ and $\mathbf{v}$ fields are represented in their three independent components, without any component truncation. Here, the electromagnetic gauge for vector potential uniqueness is set as $\Phi=0$ (Weyl gauge), thus obtaining an induction equation of the form
\begin{equation} \label{Full MHD induction eq}
   \frac{\partial \mathbf{A}}{\partial t} = -\mathbf{E} = \mathbf{ v \times B} - \eta \mathbf{J}
\end{equation}
The full MHD model renders itself necessary for the accurate reproduction of certain instabilities like internal kink modes\, \color{blue} [\onlinecite{pamela2020extended}]\, \color{blue} [\onlinecite{hoelzl2021jorek}] \color{black}. 
\subsubsection*{Coupling terms}

As anticipated in Section \ref{sec:intro} and as one can infer from Equation \eqref{SW yield}, a self-consistent electromagnetic coupling envisages the exchange of information about the vacuum and the integrated passive-active system of coils; this information is contained in $\mathbf{B \times n} \rvert_{bnd}$ and it is needed to compute the plasma equilibrium in free boundary configuration and to evolve its temporal dynamics.
The present section addresses the adopted strategies on the JOREK side of the coupling, namely, considering $\mathbf{B \times n} \rvert_{bnd}$ as an input from STARWALL. JOREK is programmed to always make use of this input to evolve $\mathbf{A}$ in time and, in return, provide $\mathbf{A \times n} \rvert_{bnd}$ to compute the $\mathbf{B \times n} \rvert_{bnd}$ at the successive time step. This $ \left[ \mathbf{A \times n} \right] \leftrightarrow  \left[ \mathbf{B \times n} \right] $ loop is entirely performed inside JOREK, which requires a set of response matrices provided by STARWALL; these are static in time and only depend on the geometry of the JOREK domain's boundary and the conductors.

    In reduced MHD, the coupling for the only component of $\mathbf{A} = \psi \nabla \phi$ is carried out leveraging the $\phi$-projection of the magnetostatic equation $ \curl \curl \mathbf{A} = \mu_0 \mathbf{J}$, which yields the form
\begin{equation} \label{Delta Star}
- \mu_0 R J_{\phi} = R^2 \nabla \cdot \left( \frac{1}{R^2} \nabla \psi \right) = R \frac{\partial}{\partial R} \left( \frac{1}{R} \frac{\partial \psi}{\partial R} \right) + \frac{\partial^2 \psi}{\partial Z^2} = \Delta^* \psi 
\end{equation}
which is expressed in a more compact form as $\Delta^* \psi = j$ according to JOREK's normalization \,\color{blue} [\onlinecite{hoelzl2021jorek}] \color{black}. In weak form, this constraint reads
\begin{equation} \label{weak form RMHD}
\iiint_{V_J} j \frac{j_l^*}{R^2} \, dV + \iiint_{V_J} \frac{1}{R^2} \nabla j_l^* \cdot  \nabla \psi \, dV - \oiint_{\partial V_J} \frac{j_l^*}{R}  \underbrace{\left( \nabla \psi \cdot \mathbf{n} / R \right)}_{B_{tan,pol}}  \, dS = 0
\end{equation}
where the scalar test functions $j_l^*$ are chosen from the same space as the JOREK basis functions, applying Galerkin's method. The tangent poloidal field $B_{tan,pol}$ is the sole component needed to perform the reduced MHD coupling and it is provided by STARWALL as a linear combination of poloidal flux on the boundary and wall currents \color{blue} [\onlinecite{hoelzl2012coupling}] \color{black}. 

\medskip
In the full MHD coupling scheme introduced in this contribution, the coupling equation is the time-varying induction equation \eqref{Full MHD induction eq}. Again, the coupling term arises from the $\eta \mathbf{J}$ component when considering its weak form, this time via dot product with a vector test function $\mathbf{F}^*$:
\begin{eqnarray} \label{FMHD coupl eq}
   && \iiint_V \eta \mathbf{F}^* \cdot \mathbf{J} \, dV = \iiint_V \eta \mathbf{F}^* \cdot (\curl \mathbf{B}) \, dV = \nonumber \\
   & & \oiint_{\partial V} \eta (\mathbf{B} \times \mathbf{F}^*) \cdot \mathbf{n} \, dS + \iiint_V \mathbf{B} \cdot \curl \left( \eta \mathbf{F}^* \right) \, dV
\end{eqnarray}
where the vector identity $\diver (\mathbf{X} \times \mathbf{Y} ) = \mathbf{Y} \cdot \curl \mathbf{X} - \mathbf{X} \cdot \curl \mathbf{Y}$ and the divergence theorem have been used.
\\
By further manipulation, $\mathbf{B \times n}$ appears in the boundary term:
\begin{equation} \label{Bxn arise}
\oiint_{\partial V} \eta (\mathbf{B} \times \mathbf{F}^*) \cdot \mathbf{n} \, dS = - \oiint_{\partial V} \eta (\mathbf{B \times n}) \cdot \mathbf{F}^* \, dS
\end{equation}
and $\mathbf{B \times n}$ is the one provided by STARWALL, by means of the response matrices.

\subsection{\label{subsec:virtual} \normalsize{Virtual casing principle}} 

\begin{figure}[h]
   \centering
   \includegraphics[width=0.4\linewidth]{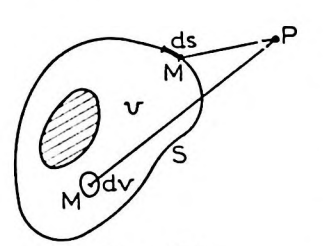}
   \caption{\small{Test volume with volumetric current distributions and field evaluation point on the outside, as a purely electromagnetic model\, \color{blue} [\onlinecite{magnetostatique}] \color{black} for the JOREK-STARWALL coupling.}}
   \label{fig:Durand}
\end{figure}

In the following, we consider the JOREK code as a black box whose output is the $\mathbf{A \times n} \rvert_{bnd}$ on its computational boundary.
With specific reference to the case being analyzed, it is possible to prove that this term contains all the information that needs to be provided from JOREK to STARWALL for a self-consistent coupling. \\
Let consider a generic volume $V$, immersed in the $\Omega_{ext}$ space, and a field evaluation point $P$ on the outside of it (fig.\@ \ref{fig:Durand}). \\
 The following can be stated:

\medskip

\underline{\textbf{Theorem 1}} \, \textit{Given the current density $\mathbf{J}_{ext}$ in the external domain $\Omega_{ext}$ and the tangent vector potential $\mathbf{n \times A} \rvert_{\partial V}$ on the boundary of the internal volume $V$, the outer magnetostatic problem}
\begin{align*}
& \curl \mathbf{B} = \mu_0 \mathbf{J} \qquad \text{in} \, \Omega_{ext} \\
& \mathbf{B} = \curl \mathbf{A} \qquad  \text{in} \, \Omega_{ext} \\
& \mathbf{n \times A}\rvert_{\partial V} = \mathbf{a}_t \quad \text{on} \, \partial \Omega_{ext}
\end{align*}
\textit{has unique solution, provided that $\mathbf{B}$ is regular at infinity.}

\medskip
\textbf{Proof}: The proof is obtained by contradiction of the hypothesis of existence of two different solutions $\mathbf{B_1}$ and $\mathbf{B_2}$. Let define the difference $\delta \mathbf{B} = \mathbf{B_1 - B_2}$. Due to the linearity of the problem, even $\delta \mathbf{B}$ has to satisfy the magnetostatic problem, which in its case has no sources $\mathbf{J}$ or $\mathbf{v}$ (associated \textit{homogeneous} problem).
\\
Let calculate the $L^2$-norm of $\delta \mathbf{B}$:
\begin{equation*}
    \lvert \lvert \delta \mathbf{B} \rvert \rvert^2_{L^2} = \iiint_{\Omega_{ext}} \lvert \delta \mathbf{B} \rvert^2 \, dV = \iiint_{{\Omega_{ext}}} \delta \mathbf{B} \cdot \nabla \times \delta \mathbf{A} \, dV
\end{equation*}
which, exploiting the vector identity $\nabla \cdot ( \mathbf{A} \times \mathbf{B} ) = \mathbf{B} \cdot \nabla \times \mathbf{A} - \mathbf{A} \cdot \nabla \times \mathbf{B}$ yields:
\begin{equation*}
    \lvert \lvert \delta \mathbf{B} \rvert \rvert^2_{L^2} = \iiint_{{\Omega_{ext}}} \delta \mathbf{A} \cdot \underbrace{\nabla \times \delta \mathbf{B}}_{=0} \, dV + \oiint_{- \partial V } \underbrace{\mathbf{n} \times \delta \mathbf{A} }_{=0} \cdot  \, \delta \mathbf{B} \, dS
\end{equation*}
It can therefore be concluded that, as long as the hypotheses of the theorem are satisfied, the $L^2$-norm of the difference magnetic field is $0$ and $\mathbf{B_1 = B_2}$, which proves that the solution to the problem is unique. \\
The main consequence of the theorem is that, as long as the evaluation point $P$ is placed on the outside of the source volume, the magnetic field generated by \underline{any} volumetric current distribution is uniquely identified by means of the tangent component $\mathbf{A} \times \mathbf{n}$ on the surface enclosing the volume, with obvious benefits in terms of computational cost.

Bearing this result in mind, let consider the JOREK computational volume as $V$. The boundary delimiting $V$ coincides with the coupling surface $S$, where we want the sources for the external magnetic field to appear. In our coupling scheme, the $\mathbf{a}_t$ term from the theorem is assigned and coincides with the vector potential $\mathbf{A}\rvert_{bnd}$. The electromagnetic problem comes down to finding the equivalent superficial sources given the vector potential produced on the boundary by the volumetric sources. \\
For this purpose, the boundary is treated as a \emph{perfectly conducting shell} from the wall code's perspective: in this scenario, any current variation appearing in the inner domain will be exactly shielded out by the equivalent surface currents induced on the shell. This implies that it is possible to equate the external magnetic field generated by internal volumetric currents to the external magnetic field generated by these so-called \emph{virtual currents} which mimic the effect of the former. 

\subsection{\normalsize{STARWALL model}} \label{subsec:SWmodel}
STARWALL solves the Maxwell MQS problem
\begin{align}
& \label{inductioneq} \mathbf{E} = -\frac{\partial \mathbf{A}}{\partial t} - \nabla \Phi \\
& \label{rotrotA} \nabla \times \nabla \times \mathbf{A} = \mu_0 \mathbf{J} \\
& \nabla \cdot \mathbf{J} = 0\\
\label{Ohm conduc} & \mathbf{E}_c = \eta_c \mathbf{J}_c 
\end{align}
where the subscript $c$ denotes conducting structures. The present system is linear, with unique solution, as long as the tangential component $\mathbf{A \times n}$ is specified on the JOREK boundary. Concerning the choice for the vector potential's uniqueness, the \emph{Coulomb gauge} $\diver \mathbf{A} = 0$ is chosen so that
\begin{equation} \label{A uniqueness}
   \curl \curl \mathbf{A} = \nabla \left( \underbrace{\diver \mathbf{A}}_{= 0} \right) - \nabla^2 \mathbf{A} = \mu_0 \mathbf{J}
\end{equation}
This allows to invert Equation \eqref{A uniqueness} via Green's functions, thus obtaining the Biot-Savart law

\begin{equation} \label{Ampère law STARWALL}
   \mathbf{A(r)} = \frac{\mu_0}{4 \pi} \iiint_{V} \frac{\mathbf{J(r^')}}{\lvert \mathbf{r - r^' \rvert}} dV^'
\end{equation}
Equation \eqref{Ampère law STARWALL} is to be interpreted as follows: the vector potential $\mathbf{A}$ in a generic point $\mathbf{r}$ is given by the right-hand-side integrals performed over all the volumes contaning source currents. In the case of our coupling, the domains where current fields appear are the conductive structures (wall and coils) and the plasma. The latter can be treated as a perfectly conducting surface as per the virtual casing principle in section \ref{subsec:virtual}. STARWALL implements a FEM discretization of conducting structures via triangular elements; instead, the vacuum region in which the plasma and the conducting structures are immersed is not discretized, leveraging exactly the virtual casing principle. The code relies on a \emph{thin wall approximation}, which envisages that the adopted representation for the conductors is one of infinitely thin structures. The two main assumptions are:
\begin{itemize}
   \item Wall thickness is much smaller than its characteristic length.
   \item Electric currents are homogeneous in the direction perpendicular to the conducting thin surfaces.
\end{itemize}
The thin wall approximation is valid when the diffusion of currents across the thickness of the conducting structure is much faster then the dynamics of the instability under investigation. If the inducing magnetic field varies on a time-scale $\tau$, it penetrates into the conductor in the limits dictated by the skin depth $\delta_{skin} = \sqrt{\eta \tau / (\pi \mu_0)}$. If the conductor's thickness is much smaller then the skin depth, the current can be deemed homogeneous across the conductor; this implies that, in order for the thin wall approximation to be valid, the time-scale of the instability is required to \mbox{satisfy} the condition
\begin{equation} \label{thinwallcond}
\tau >> \frac{\pi \mu_0 d_w^2}{\eta_w}
\end{equation}
where $\eta_w$ is the wall resistivity and $d_w$ is the thickness. 
This requirement is satisfied for the classes of instabilities being reproduced in the following sections of this contribution. 

\medskip

Combining the virtual casing principle and the thin wall approximation, the STARWALL model ends up being comprised of two magnetically coupled surfaces (virtual shell and thin wall) interacting via changes in the magnetic field. \\
Therefore, the integrals in equation \eqref{Ampère law STARWALL} are simplified to surface integrals, with obvious numerical advantages: 
\begin{equation} \label{surfaceapprox}
\mathbf{A}(\mathbf{r}) = \frac{\mu_0}{4 \pi} \iiint_{V} \frac{\mathbf{J}(\mathbf{r^'})}{\lvert \mathbf{r}-\mathbf{r^'} \rvert} \, dV^' = \frac{\mu_0}{4 \pi} \iint_{S} \frac{\mathbf{j}(\mathbf{r^'})}{\lvert \mathbf{r}-\mathbf{r^'} \rvert} \, dS^'
\end{equation}
Plugging equation \eqref{surfaceapprox} into the induction equation \eqref{inductioneq} one gets to the form

\begin{equation} \label{eddy currents formulation}
   \frac{\partial \mathbf{A}}{\partial t} = - (\mathbf{E} + \nabla \Phi ) = \frac{\mu_0}{4 \pi} \iiint_S \frac{\partial \mathbf{j(r^')}}{\partial t} \frac{1}{\lvert \mathbf{r - r^'} \rvert} \, dS^'
\end{equation}

where the $\mathbf{j}$ currents are surface currents flowing in the two surfaces at play.
The two equations \eqref{surfaceapprox} and \eqref{eddy currents formulation} are applied to the virtual shell and the thin wall respectively, so as to impose the necessary boundary conditions to close the system of equations:
\begin{align}
   \label{Ap equation} \mathbf{A}_p & = \frac{\mu_0}{4 \pi} \iint_S \frac{\mathbf{j(r^')}}{\lvert \mathbf{r - r^'} \rvert} \, dS^' \\
   \label{jw, phiw equation} -\frac{\eta_w \mathbf{j}_w}{d_w} - \nabla \Phi_w & = \frac{\mu_0}{4 \pi} \iint_S \frac{\partial \mathbf{j(r^')}}{\partial t} \frac{1}{\lvert \mathbf{r - r^'} \rvert} \, dS^'
\end{align}
where in equation \eqref{Ap equation} the vector potential on the coupling surface is the input from JOREK, and in equation \eqref{jw, phiw equation} the electric field in the conducting structures is given by Ohm's law. This set of equations completely describes the eddy current model and brings up the related response matrices when a solution via finite elements is applied.
The STARWALL test function, which by definition must share the same properties as the interpolated function (i.\@e.\@ the surface current density $\mathbf{j}$) is chosen as
\begin{equation} \label{test func j}
   \mathbf{j}^* = \nabla I^* \times \mathbf{n}
\end{equation}
where the $I$ functions are \emph{current potentials} and $\mathbf{n}$ is the normal unit vector to the coupling surface or to the wall + coils surface, depending on where the test function is defined. Note that $\mathbf{j}^*$ satisfies the divergence-free condition $\diver \mathbf{j}^* = 0$ by construction.
The weak form of the eddy current model is now obtained by performing the scalar product with test functions and integrating over the corresponding surface for each equation:
\begin{equation}
\label{Ap FEM} \iint_{S_p} \mathbf{j}_p^* \cdot \mathbf{A}_p \, dS_p = \frac{\mu_0}{4 \pi} \sum_{i=1}^{N_{\Delta}} \iint_{S} \; \iint_{S_p} \frac{\mathbf{j}_p^* \cdot \mathbf{j}_i(\mathbf{r^'})}{\lvert \mathbf{r - r^'} \rvert} \, dS_i^' \, dS_p \\
\end{equation}
\begin{eqnarray} \label{jw, phiw FEM} 
- \iint_{S_w} \frac{\eta_w \mathbf{j}_w^* \cdot \mathbf{j}_w}{d_w} \, dS_w - \oint_{\partial S_w} \Phi_w \nabla I_w^* \cdot \, d\mathbf{l}_w = \nonumber \\
= \frac{\mu_0}{4 \pi} \sum_{i=1}^{N_{\Delta}} \iint_{S} \; \iint_{S_w} \frac{\mathbf{j}_w^* \cdot \frac{\partial \mathbf{j}_i(\mathbf{r^'})}{\partial t}}{\lvert \mathbf{r - r^'} \rvert} \, dS_i^' \, dS_w  
\end{eqnarray}
where the sum is performed over the $N_{\Delta}$ STARWALL triangular elements. The current potentials  introduced in (\ref{test func j}) are described via linear interpolation in each triangular element. Two types of current potentials are necessary to describe the effective current fields' structure on the two surfaces: in fact, when dealing with toroidally and poloidally closed paths, \emph{net current potentials} $I^T, I^P$ have to be added to the \emph{single-valued potentials} $I_s$ describing the eddy currents. The reason for the introduction of these extra potentials is that an integral over a closed loop of a function of type $\nabla I_s$ vanishes when $I_s$ is periodic; therefore the net current potentials allow to introduce the net magnetic flux contribution for the axisymmetric mode numbers $n=0$ and $m=0$ \, \color{blue} [\onlinecite{such2018free}]\color{black}\, \color{blue} [\onlinecite{merkel2015linear}]\color{black} . While the toroidal net current potential is already present for the reduced MHD coupling, the full MHD coupling presented in this contribution has been carried out without the introduction of a net poloidal current potential, as this is not crucial for the main instabilities of interest and will be implemented in a later stage. \\
Ultimately, the STARWALL representation of eddy currents is one of divergence-free current loops sustained by current potentials as the degrees of freedom of the problem and assumed to be constant over each triangle:
\begin{equation} \label{j delta}
   \mathbf{j}_{\Delta} = \frac{I_a \mathbf{r}_{cb} + I_b \mathbf{r}_{ac} + I_c \mathbf{r}_{ba}}{2S_{\Delta}}
\end{equation}
where $S_{\Delta} = \lvert \mathbf{r}_{ba} \times \mathbf{r}_{ac} \rvert / 2$ is the triangle area, $a,b,c$ are the local triangle nodes and $\mathbf{r}_{ij} = \mathbf{r}_i - \mathbf{r}_j$.
With this representation, equations \eqref{Ap FEM} and \eqref{jw, phiw FEM} of the eddy current model are brought into the FEM matrix form
\begin{align}
  \label{Ap matrix form} & \underline{M}_{pp} \mathbf{I}^p + \underline{M}_{pw} \mathbf{I}^w = \sum_{l=1}^{2} \underline{M}_{pe,l} \mathbf{A}^e_{p,l}/ \mu_0 \\
   \label{jw, phiw matrix form} & \underline{M}_{wp} \dot{\mathbf{I}}^p + \underline{M}_{ww} \dot{\mathbf{I}}^w = -\underline{R}_{ww} \mathbf{I}^w / \mu_0 
\end{align}
Here, the $\mathbf{I}^p$ and $\mathbf{I}^w$ vectors contain the STARWALL degrees of freedom for the virtual shell and the wall surface currents respectively, in terms of the potentials introduced in equation \eqref{test func j}, while $\mathbf{A}^e_{p,l}$ contains the virtual shell tangent vector potential degrees of freedom, with $l$ distinguishing between the toroidal and poloidal component. \\
The response matrices are computed by STARWALL and, after further adaptation and projection into the Bézier space, they are passed on to JOREK, which solves the evolution of wall currents at each time step of the simulation. A glossary of the STARWALL matrices can be found in Appendix \ref{Matrix glossary}.

\section{\normalsize{Full MHD coupling}} \label{Coupling scheme}
Extending the electromagnetic coupling to a full MHD configuration entails the complete representation of the $\mathbf{B \times n}$ component on the STARWALL side, in order to inform the JOREK coupling term from equation \eqref{Bxn arise} with the suitable natural boundary condition. The $\mathbf{B}_{tan}$ equation is provided by applying the $\curl$ operator to the Biot-Savart law with the STARWALL discretization as presented in \ref{subsec:SWmodel}. 
\begin{equation}
\begin{split}
     \mathbf{B}_{tan} \rvert_{JOR} & = \frac{\mu_0}{4 \pi} \curl \left( \sum_{i=1}^{N_{\Delta}} \mathbf{j}_{\Delta_i} \iint_{\Delta_i} \frac{1}{\lvert \mathbf{r-r^'} \rvert} \, dS^' \right) \times \mathbf{n} \\
     & = \frac{\mu_0}{4 \pi} \mathbf{n} \times \left( \sum_{i=1}^{N_{\Delta}} \mathbf{j}_{\Delta_i} \times \nabla \iint_{\Delta_i} \frac{1}{\lvert \mathbf{r-r^'} \rvert} \, dS^' \right)
\end{split}
\end{equation}
and performing these integrals at the JOREK boundary\footnote{The $\nabla \left( \frac{1}{\lvert \mathbf{r-r^'} \rvert} \right)$ function incurs a singularity when the field points $\mathbf{r}$ approach the source points $\mathbf{r^'}$, which is what happens on the JOREK boundary. The numerical solution to this issue is to shift the field points on an auxiliary surface $S_p^'$ displaced radially by a $\delta$ quantity with respect to the JOREK boundary $S_p$; therefore the field points are at $\mathbf{r = r^'+ n}\delta$ and this surface is called \emph{control surface}.} and at the wall surface results in an expression for the matrix form of $\mathbf{B}_{tan}$:
\begin{equation} \label{Btan first form}
\mathbf{B}_{tan}/\mu_0 = \underline{\mathbf{M}}_{\parallel w} \mathbf{I}^w + \underline{\mathbf{M}}_{\parallel p} \mathbf{I}^p    
\end{equation}
where the bold symbol $\underline{\mathbf{M}}$ means that the matrix is component-wise, that is, it accounts for both the poloidal and toroidal tangent fields.
\\
Expressing $\mathbf{I}^p$ as a function of $\mathbf{I}^w$ as per equation (\ref{Ap matrix form}) yields:
\begin{equation} \label{Btan second form with A}
   \mathbf{B}_{tan}/\mu_0 = \hat{\underline{\mathbf{M}}}_{\parallel w} \mathbf{I}^w + \sum_{l=1}^3 \underline{\mathbf{M}}_{\parallel e,l} \mathbf{A}_{p,l}^e / \mu_0
\end{equation}
where
\begin{align}
   \hat{\underline{\mathbf{M}}}_{\parallel w} & \equiv \underline{\mathbf{M}}_{\parallel w} - \underline{\mathbf{M}}_{\parallel p} \underline{M}_{pp}^{-1} \underline{M}_{pw} \\
   \underline{\mathbf{M}}_{\parallel e,l} & \equiv \underline{\mathbf{M}}_{\parallel p} \underline{M}_{pp}^{-1} \underline{M}_{pe,l}
\end{align}
and these matrices model the contributions of the plasma tangent vector potential and the wall + coils current potentials on the tangential field (Appendix \ref{Matrix glossary}). \\
The \emph{vacuum response matrix} $\underline{\mathbf{M}}_{\parallel e,l}$ is the one realizing the link $\mathbf{B \times n} \leftrightarrow \mathbf{A \times n}$: it is noteworthy that, in the reduced MHD coupling scheme, this matrix was only expressing the coupling between $\mathbf{B}_{tan,pol}$ and $\mathbf{A}_{tan,tor}$, which were the only components being evolved by the natural BC; in the full MHD scheme the matrix is fourfold and it models the interactions $\mathbf{B}_{tan,pol} \leftrightarrow \mathbf{A}_{tan,tor}$,  $\mathbf{B}_{tan,pol} \leftrightarrow \mathbf{A}_{tan,pol}$,$\mathbf{B}_{tan,tor} \leftrightarrow \mathbf{A}_{tan,tor}$,  $\mathbf{B}_{tan,tor} \leftrightarrow \mathbf{A}_{tan,pol}$. From a physical standpoint, the relaxation of the complete set of degrees of freedom for the $\mathbf{A}$ and $\mathbf{B}$ fields allows for a more accurate reproduction of plasma dynamics, especially at harmonics $n>0$, for which the self-interactions $\mathbf{B}_{tan,pol} \leftrightarrow \mathbf{A}_{tan,pol}$,$\mathbf{B}_{tan,tor} \leftrightarrow \mathbf{A}_{tan,tor}$ were being completely neglected in the reduced MHD coupling scheme. \\
As it is done for reduced MHD, the set of full MHD response matrices undergoes a transformation to Bézier space and is used by JOREK for the computation of the wall currents and the evolution of the boundary vector potential at each time step. \\
As already pointed out in Section \ref{JOREKside}, the main difference on the JOREK side with respect to reduced MHD is the change of the coupling equation, which becomes time-varying in the full MHD scheme.

\section{\normalsize{Vacuum response tests}} \label{Tilted wire}
\begin{figure} [h] 
   \includegraphics[width=0.8\linewidth]{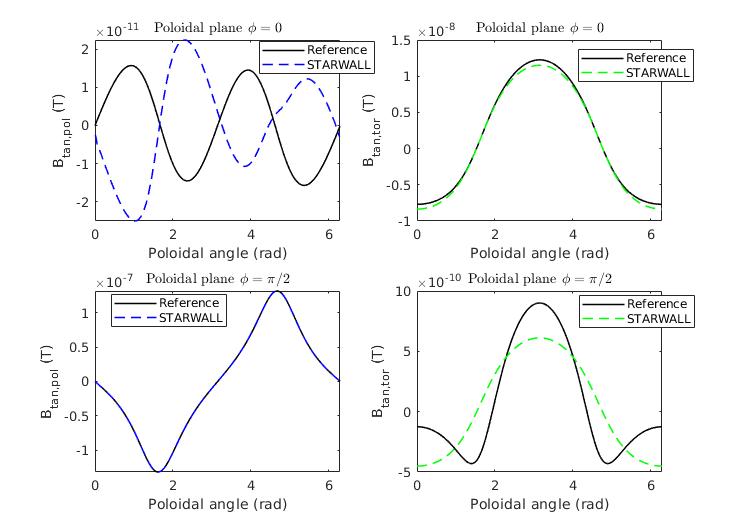}
   \caption{\small {Comparison of reference fields (black) and STARWALL-computed fields at $\phi = 0$ and $\phi = \pi / 2$ , both poloidal (blue) and toroidal (green) for a $n_{pol} = 900, n_v = 120 $ discretization. For the chosen configuration (3° tilted wire), $B_{pol}$ has a predominant sine component and $B_{tor}$ has a predominant cosine component. Because of resolution limitations, the lower-amplitude harmonic components (top left and bottom right) do not match well, but the effect on the global error is negligible. }}
   \label{fig:STWfields}
\end{figure}

The most straightforward method to assess the consistency of the full MHD STARWALL vacuum response consists in generating a test vector potential on the JOREK boundary and feeding it to the $\mathbf{M}_{\parallel e,l}$ matrices to obtain the resulting tangent $\mathbf{B}$ and then verify it against analytical solutions. Equivalently, this test verifies whether the shell virtual currents $\mathbf{j}_p$ can faithfully reproduce the magnetic field of whatever current distribution in the bulk of the JOREK domain. A case in which the analytical solution of a magnetic field with multiple toroidal harmonics (needed to test the response at higher modes) is known is one of a tilted toroidal wire placed at the center of the JOREK domain. Note that no wall is present in this scenario, as the sole vacuum response is being tested. \\
The simulations have been run for the $n=1$ field harmonic, where each harmonic has the expression

$$ B_n^{c} (R,Z) cos(n \phi) + B_n^{s}(R,Z) sin (n \phi) $$
as they are extracted from a spectral expansion. Figure \ref{fig:STWfields} displays the toroidal and poloidal fields' cosine and sine components given by the above expression (i.\@e.\@ in real space) by fixing the toroidal angles at $\phi = 0$ and $\phi = \pi / 2$ respectively.

The discretization is handled as follows:
\begin{itemize}
\item The JOREK boundary is axisymmetric and comprised of $n_{bnd}$ 1D elements;

\newpage

\item The user can choose to further subdivide\footnote{A subdivision is anyway needed because the coupling requires lower order finite elements, compared to JOREK's higher-order FEM \color{blue} [\onlinecite{czarny2008bezier}] \color{black}.} each boundary element into $n_{sub,\theta}$ poloidal sub-elements, therefore the total number of elements determining the poloidal resolution is $n_{pol} = n_{bnd} * n_{sub,\theta} $;
\item The number of points in the toroidal direction can be chosen as well ($n_{v}$).
\end{itemize}

A first test has been run with fixed $n_{bnd}=6, n_{sub, \theta} = 10$ and variable toroidal resolution. The root relative square error has been evaluated for the harmonic poloidal and toroidal components, according to the definition
\begin{equation}
RRSE_k = \frac{\sqrt{\frac{1}{N_{bnd}} \sum_{i=1}^N  \left[ B_{ref,k}(R_i,Z_i)-B_k(R_i,Z_i) \right]^2}} {\sqrt{\frac{1}{N_{bnd}} \sum_{i=1}^N B_{ref,k}(R_i,Z_i)^2 }}
\end{equation}

\begin{figure}[h]
   \centering
   \includegraphics[width=0.7\linewidth]{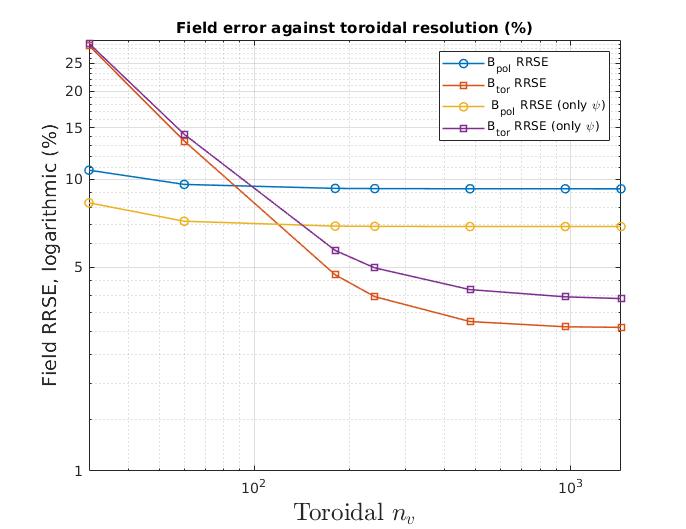}
   \caption{\small{Toroidal resolution scan (with $n_{pol} = 60$) of the RRSE for the $n=1$ components of poloidal and toroidal field generated by a 3° tilted toroidal wire, logarithmic scale.}}
   \label{fig:nv trend}
\end{figure}

The RRSE with respect to the Biot-Savart analytical solution displays a consistent reduction trend against the increase of toroidal resolution (figure \ref{fig:nv trend}). The benefit attained from the relaxation of the $A_{pol}$ degrees of freedom on the JOREK boundary can also be appreciated, as the toroidal field error is lower compared to a $\psi$-only coupling scheme like the one implemented in reduced MHD. Due to the low poloidal resolution chosen for this case, this improvement cannot be yet observed for the $B_{pol}$ error, which saturates at around $10^2$ toroidal points.

\begin{figure}[h]
   \centering
   \includegraphics[width=0.7\linewidth]{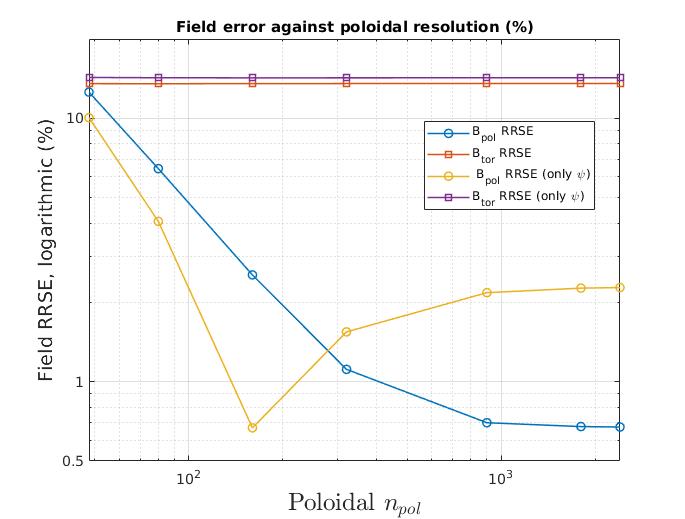}
   \caption{\small{Poloidal resolution scan (with $n_v = 60$) of the RRSE for the $n=1$ components of poloidal and toroidal field generated by a 3° tilted toroidal wire.}}
   \label{fig:npol trend}
\end{figure}

\medskip
A scan on the poloidal resolution has also been performed (figure \ref{fig:npol trend}) by fixing $n_v = 60$. In this case, the toroidal field error is already converged and it is independent of the poloidal resolution; the error on the full MHD poloidal field $B_{pol}(\psi, A_{pol})$ displays a consistent reduction trend, while the error trend for the reduced-MHD poloidal field  $B_{pol}(\psi)$ is revealed to be similar to vacuum tests performed for the CARIDDI reduced MHD coupling\, \color{blue} [\onlinecite{isernia2023self}]\color{black}.

The resolution scans show that the representation of the magnetic field components can be made well faithful to realistic and production cases by choosing a suitable discretization. \\
In conclusion:
\begin{itemize}
\item For the particular test case (3° tilted toroidal wire) the tangent toroidal field component is 1 order of magnitude smaller compared to the poloidal component; this results in a toroidal field error about 1 order of magnitude larger.
\item In order to achieve the same degree of accuracy in both components, the resolution requirements increase significantly, since way more toroidal grid points are now needed compared to reduced MHD cases.
\end{itemize}

\section{\normalsize{Validation of the Full MHD coupling}} \label{Simulations}
The present section is aimed at showcasing the simulation of the full MHD coupling scheme for different families of MHD modes, with the scope of assessing its consistency and observing the differences with respect to the JOREK-STARWALL reduced MHD coupling scheme (internal benchmarks) and other full MHD codes (external benchmarks). In particular, both $n=0$ axisymmetric VDEs and $n=1$ non-axisymmetric tearing modes have been reproduced with the new model.
\subsection{\normalsize{Axisymmetric VDE test case}} \label{VDE}

\begin{figure}[h]
   \centering
   \includegraphics[width=0.8\linewidth]{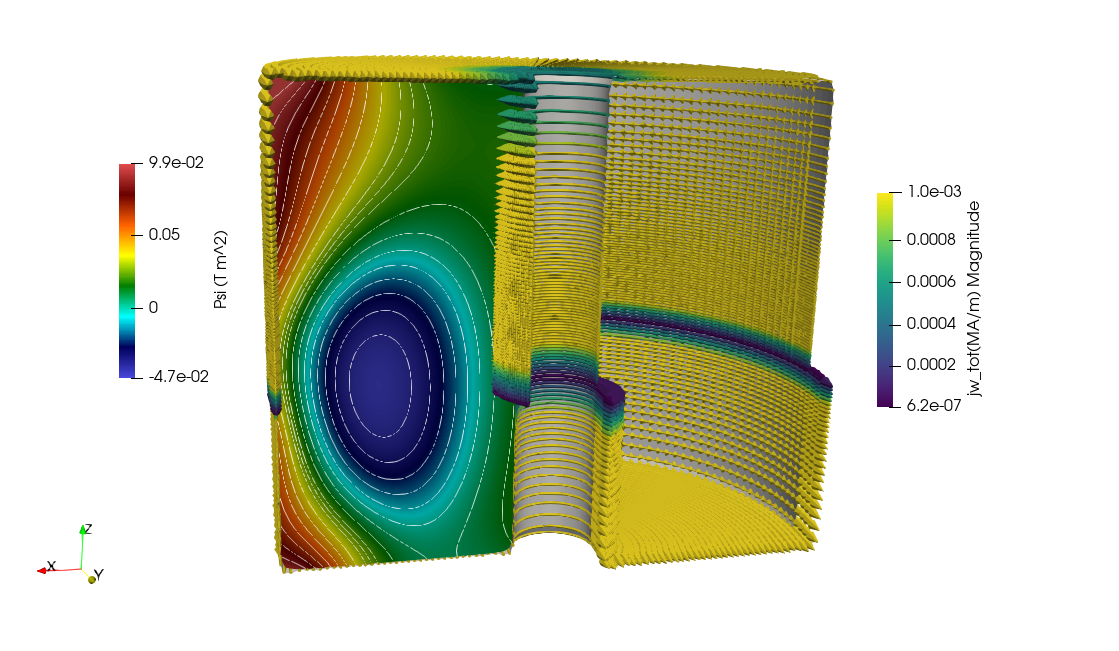} \caption{\small{Snapshot of the plasma poloidal flux surfaces during an axisymmetric vertical displacement event in the NSTX machine. The induced up-down antisymmetric eddy current field in the wall is also visible.}}
   \label{fig:VDE screen}
\end{figure}

In the following, the simulation of a realistic hot Vertical Displacement Event (VDE) in the NSTX tokamak is presented for the reduced and the full MHD models of JOREK (figure \ref{fig:VDE screen}).\\
In particular, the same case as \color{blue} [\onlinecite{krebs2020axisymmetric}]\color{black} \, is considered. The non-linear simulations already performed with JOREK (reduced), NIMROD and M3D-C$^1$ (full) showcase an agreement between the full and reduced MHD models that is around 3\% or less for most wall resistivities, and is rarely around 10\%. 
\\
In this contribution, we aim to reproduce the same wall resistivity scan using the novel JOREK full MHD model. The sole difference with respect to \color{blue} [\onlinecite{krebs2020axisymmetric}]\color{black} \, is that we introduce no offset temperature $T_{e,\text{off}}$ at the edge, which will result in a slowdown of the plasma displacement due to the dampening action of response currents in the open field lines region. \\
During the vertical displacement of a plasma with elevated core temperature, the current diffusion time happens to be longer than the vertical drift time\, \color{blue}[\onlinecite{artola2024modelling}] \color{black}. As a result, the plasma current $I_p$ is roughly conserved in the initial phase of the instability (before the current quench) as the currents flowing outside of the Last Closed Flux Surface (LCFS) decay, and are therefore induced back in the core plasma. In terms of vector potential, this entails a small contribution from the $\mathbf{A}_{pol}$ dynamics modeled by full MHD, as $A_{\phi} >> A_{pol}$ during the displacement. In light of this observation, we here present the VDE simulation using a full MHD model where only the $A_{\phi}$  dynamics is active on the boundary, and the poloidal resistivity $\eta_{pol}$ is set to be much smaller than the one in toroidal direction $\eta_{\phi}$. By doing so, the poloidal term in equation \eqref{FMHD coupl eq} is shut off, but the rest of the full MHD physics is retained. \\
\begin{figure}[h]
   \centering
   \includegraphics[width=0.6\linewidth]{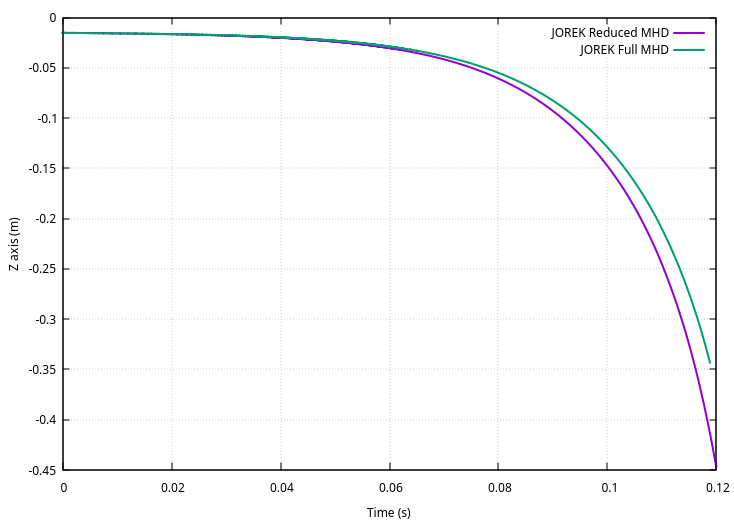} \caption{\small{Time evolution of the magnetic axis' vertical position during a VDE in NSTX, simulated with JOREK-STARWALL in reduced and full MHD. The thin wall resistivity is set to $ \eta_w / d_w = 200 \times 10^{-6} \; \Omega $. }}
   \label{fig:VDE comp}
\end{figure}

\color{black} 

\begin{figure}[h]
   \centering
   \includegraphics[width=0.6\linewidth]{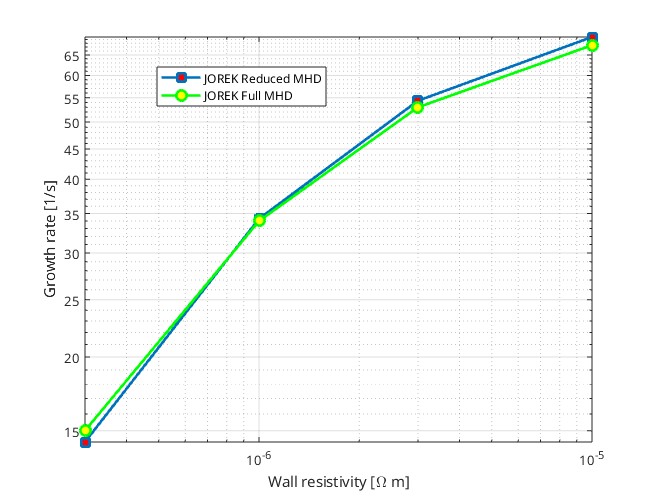} \caption{\small{Wall resistivity scan for the NSTX $n=0$ VDE simulated by reduced and full MHD JOREK.}}
   \label{fig:VDE bench}
\end{figure}

For this test, the STARWALL discretization for the virtual coupling shell has been set to $n_{pol} = 2000$, $n_v = 52$, while for the wall $n_{w,u} = 250$ points in poloidal direction and $n_{w,v} = 150$ points in toroidal direction have been used.
For each wall resistivity, an exponential fitting of type $Z_{\text{axis}} = a + b \cdot \text{exp}(\gamma \, t)$ has been adopted over the linear phase of the simulation, for the evaluation of the growth rate $\gamma$ (fig.\@ \ref{fig:VDE comp}).\\
As displayed in figure \ref{fig:VDE bench}, the growth rate agreement between Reduced and Full MHD always stays below 5\%, in agreement with the previous $n=0$ non-linear benchmark.

\color{black}
\subsection{\normalsize{Tearing Mode test case}}

\begin{figure}[h]
   \centering
   \includegraphics[width=0.7\linewidth]{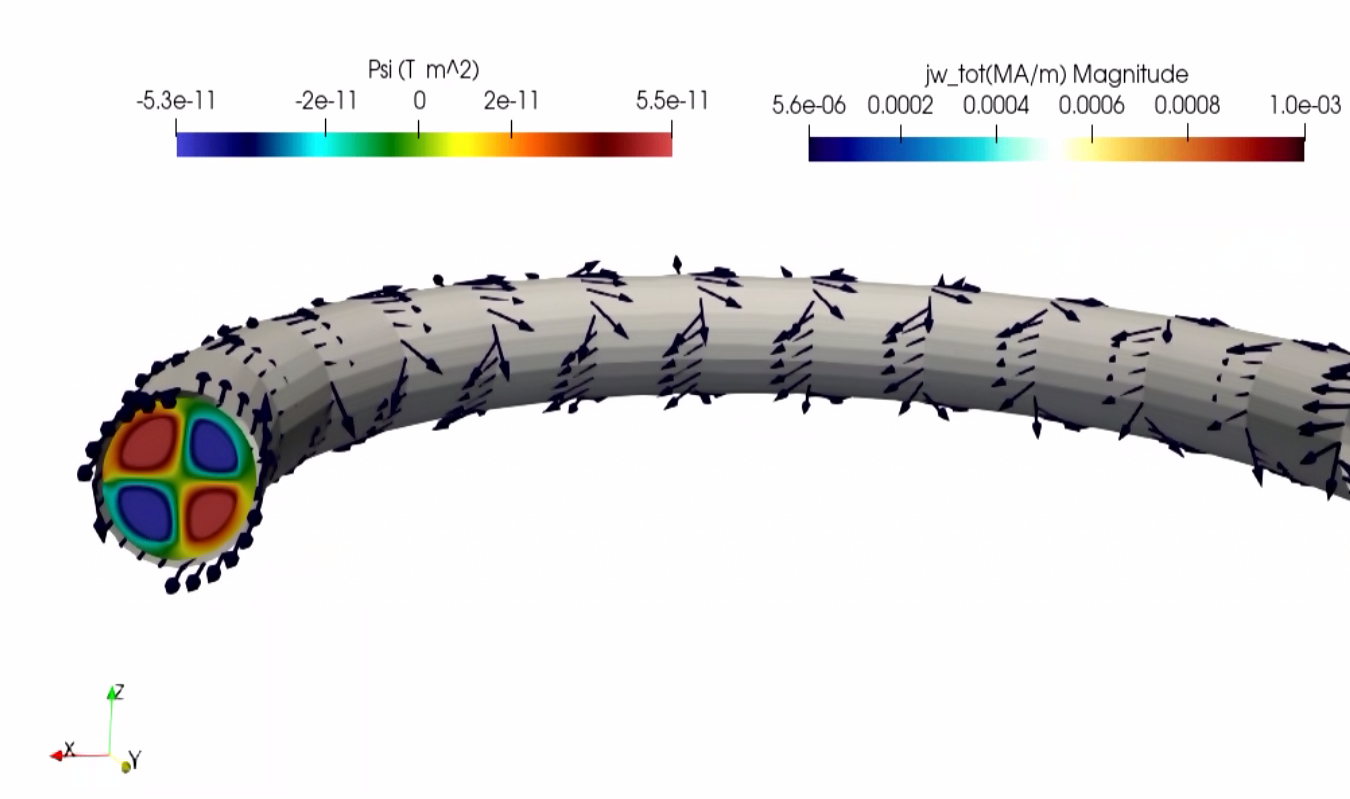}
   \caption{\small{Reproduction of $(2/1)$ tearing mode in a circular high-aspect-ratio plasma. The set of induced rotating eddy currents is visible in the wall.}}
   \label{fig:TM screen}
\end{figure}

\begin{figure}[h]
   \centering
   \includegraphics[width=0.6\linewidth]{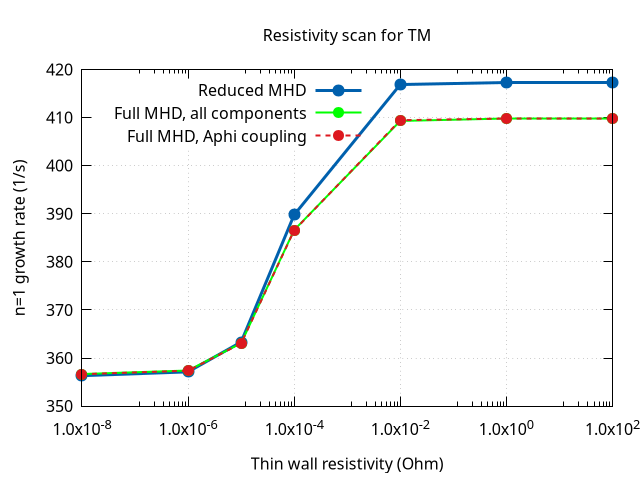}
   \caption{\small{$n=1$ tearing mode, wall resistivity scan performed with reduced and full MHD JOREK-STARWALL.}}
   \label{fig:TM scan}
\end{figure}

In this section, the full MHD coupling scheme is validated for non-axisymmetric harmonics against the reduced MHD scheme. In particular, a 2/1 tearing mode in a large aspect-ratio plasma (major radius $R = 10 \; m$, minor radius $a = 1 \; m$) is reproduced with JOREK-STARWALL (figure \ref{fig:TM screen}). All the new components of $\mathbf{A}$ are let free on the boundary in the full MHD simulation. The $q=2$ rational surface at equilibrium is chosen to be close enough to the wall, such that the effects of the interaction become more significant during the mode evolution. When a MHD resistive wall mode with a certain toroidal periodicity is excited, the induced wall eddy current field will mirror the plasma mode structure: hence, a set of toroidally "rotating" eddy currents is expected to appear. \\
The coupling surface is discretized by STARWALL with $n_{pol} = 900$ , $n_v = 40$, while the wall discretization employs $n_{w,u} = 24$ points in poloidal direction and  $n_{w,v} = 24$ points in toroidal direction. \\
According to previously made benchmarks\, \color{blue} [\onlinecite{hoelzl2012coupling}]\color{black} \, of reduced MHD JOREK against full MHD codes, the estimation of tearing modes' growth rates is in very good agreement between full and reduced MHD models. \\
In order to assess these results, a scan over the wall resistivity has been performed for the two models' growth rates using the newly implemented JOREK-STARWALL full MHD scheme. As shown in figure \ref{fig:TM scan}, a saturation can be observed for the no-wall and ideal wall limits, and the agreement is consistent with previous literature's findings.

\color{black}

In particular, the same mode has been reproduced employing both a full MHD model with complete free boundary conditions for $\mathbf{A}$ (i.\@e.\@ where $A_R$, $A_Z$ and $A_{\phi}$ evolve on the JOREK boundary) and a full MHD model with $A_{\phi}$-coupling, like the one displayed in section \ref{VDE}. \\
As it can be observed, the extension of free boundary conditions to $A_R$ and $A_Z$ has little influence on the growth rate, whose difference compared to reduced MHD is mainly attributable to the difference in plasma models itself (i.\@e.\@ the increase of degrees of freedom in $\mathbf{v}$ and $\mathbf{B}$ in the full MHD model).

\section{\normalsize{Conclusions and future work}} \label{Conclusions}
In this manuscript, we have presented the implementation and validation of the self-consistent coupling between the full MHD model of JOREK\, \color{blue} [\onlinecite{hoelzl2021jorek}]\color{black} \, and the STARWALL\, \color{blue} [\onlinecite{hoelzl2012coupling}]\color{black} \, code. The developed code suit represents the extension of the already implemented JOREK-STARWALL reduced MHD coupling. The limitations of the full MHD JOREK-STARWALL model stem from the representation of wall structures, which can only be modeled as thin conductors, and from the missing coupling of shared (halo) currents. Even though the first approximation proves to be valid for most of the plasma instabilities of interest, it can be removed by adapting the CARIDDI code\, \color{blue} [\onlinecite{isernia2023self}]\color{black} \, to the JOREK full MHD natural boundary conditions developed in this contribution. The full MHD coupling with CARIDDI is in fact a straightforward task at this stage and it is currently being prepared. Regarding the second limitation, a general halo current coupling scheme is also in preparation for JOREK-CARIDDI, as this becomes essential when dealing with the prediction of the rotating sideways forces associated with certain MHD modes and possibly causing severe damage to the machine's mechanical structure. \\
In Section \ref{Models} the JOREK and STARWALL codes have been presented, together with the methodology adopted for the coupling. \\
Section \ref{Coupling scheme} showcases the full MHD extension and the advanced modeling features of the new coupling scheme. \\
In Section \ref{Tilted wire} we display the consistency tests made on the STARWALL full MHD magnetic field: the resolution scans on the JOREK boundary reveal a consistent reduction trend of the field error with respect to analytical references. As expected, the reduction of the poloidal field error strongly depends on the poloidal resolution, and the same applies for the toroidal field error and the respective toroidal resolution. \\
The first full MHD JOREK-STARWALL simulations are finally presented in Section \ref{Simulations}. \\
The axisymmetric VDE internal benchmark exhibits a consistent behavior with respect to previously made reduced-full MHD comparisons of non-linear axisymmetric MHD codes\, \color{blue} [\onlinecite{krebs2020axisymmetric}]\color{black}. \\
Non-axisymmetric modes have also been reproduced and the match between the reduced and full MHD growth rates, observed while varying the wall resistivity, is in agreement with the literature\, \color{blue} [\onlinecite{hoelzl2012coupling}]\color{black}. 

\medskip
These first results ensure that the new full MHD free boundary model of JOREK can be used for production cases where the reduced MHD limitations need to be removed. \\
It is therefore possible to reproduce the whole picture of MHD modes, and the related wall currents and forces, with greater fidelity. 

\newpage 

\section{Acknowledgements}

The authors wish to sincerely thank Professor G.T.A. Huijsmans, Dr. N. Isernia, Professor G. Rubinacci, Dr. N. Schwarz and Professor F. Villone for their insight and fruitful discussions. \\
This work has been carried out within the framework of the
EUROfusion Consortium, funded by the European Union via the
Euratom Research and Training Programme (Grant Agreement No.
101052200–EUROfusion). Views and opinions expressed are
however those of the author(s) only and do not necessarily reflect
those of the European Union or the European Commission. Neither
the European Union nor the European Commission can be held
responsible for them. \\
ITER is the Nuclear Facility INB No. 174. This paper explores
physics processes during the plasma operation of the tokamak when
disruptions take place; nevertheless, the nuclear operator is not
constrained by the results presented here. The views and opinions
expressed herein do not necessarily reflect those of the ITER
Organization.

\section*{\normalsize Data availability}
The data underlying this study are available from the authors upon reasonable request.

\section*{\normalsize References}
\nocite{*}
\bibliography{aipsamp.bib}

\medskip 

\pagebreak

\appendix

\section{Glossary of STARWALL matrices} \label{Matrix glossary}

\subsection*{Eddy current matrices (equations \ref{Ap matrix form}, \ref{jw, phiw matrix form})}
\begin{center}
\begin{tabular}{ | P{3em} | P{3cm}| P{3.3cm} | } 
  \hline
  \textbf{Matrix} &  \textbf{Denomination} & \textbf{Definition} \\
  \hline
  $\underline{M}_{pp}$ & Plasma-plasma interaction matrix &  Self-inductance\\
  \hline
  $\underline{M}_{pw}$ & Plasma-wall interaction matrix &  Mutual inductance \\
  \hline
  $\underline{\mathbf{M}}_{pe}$ & Plasma-vacuum interaction matrix & Boundary condition, $\theta$ and $\phi$ components of $\mathbf{A}_{tan}$  \\
  \hline
  $\underline{M}_{wp}$ & Wall-plasma interaction matrix & Mutual inductance \\
  \hline
  $\underline{M}_{ww}$ & Wall-wall interaction matrix & Self-inductance  \\
  \hline
   $\underline{\hat{M}}_{ww}$ & Modified wall-wall interaction matrix & $ \underline{M}_{ww} - \underline{M}_{wp} \underline{M}^{-1}_{pp} \underline{M}_{pw} $ \\
  \hline
  $\Tilde{\underline{\mathbf{M}}}_{we}$ & Wall-vacuum interaction matrix & $ \underline{M}_{wp} \underline{M}^{-1}_{pp} \underline{\mathbf{M}}_{pe} $ \\
  \hline
  $\underline{R}_{ww}$ & Wall resistance matrix & Resistance \\
  \hline
    \end{tabular}
\end{center}

\subsection*{Field matrices (equation \ref{Btan second form with A})}
\begin{center}
\begin{tabular}{ | P{3em} | P{3cm}| P{3.3cm} | } 
\hline
\textbf{Matrix} &  \textbf{Denomination} & \textbf{Definition} \\
\hline
$\underline{\mathbf{M}}_{\parallel w}$ & Vacuum-wall interaction matrix & Boundary condition, toroidal and poloidal $\mathbf{B}_{tan}$\\
\hline
$\underline{\hat{\mathbf{M}}}_{\parallel w}$ & Modified vacuum-wall interaction matrix &  $ \underline{\mathbf{M}}_{\parallel w} - \underline{\mathbf{M}}_{\parallel p} \underline{M}^{-1}_{pp} \underline{M}_{pw} $ \\
\hline
$\underline{\mathbf{M}}_{\parallel p}$ & Vacuum-plasma interaction matrix & Boundary condition, toroidal and poloidal $\mathbf{B}_{tan}$ \\
\hline
$\underline{\mathbf{M}}_{\parallel e}$ & Vacuum response matrix & $\mathbf{B}_{tan} \leftrightarrow \mathbf{A}_{tan}$ relation $\underline{\mathbf{M}}_{\parallel e} = \underline{\mathbf{M}}_{\parallel p} \underline{M}^{-1}_{pp} \underline{\mathbf{M}}_{pe} $ \\
\hline
\end{tabular}
\end{center}

\end{document}